\documentclass[]{article}
\usepackage{lscape}
\usepackage{pdflscape}
\usepackage{graphicx}
\usepackage{multirow}
\usepackage{dcolumn}
\usepackage{natbib}
\usepackage{amsmath}
\usepackage{amssymb}
\usepackage{listings}
\usepackage[T1]{fontenc}
\usepackage[utf8]{inputenc}
\usepackage{graphicx}
\usepackage{epstopdf}
\usepackage{color}
\usepackage{float}
\usepackage{tikz}
\usepackage{tikz-cd}
\usepackage{tcolorbox}
\usepackage{multirow}
\usepackage{hhline}
\usepackage{xcolor}
\usepackage{colortbl}
\usepackage{subcaption}
\usepackage{booktabs}
\usepackage{lscape}
\usepackage{pgfplots}
\usepackage{lineno}
\usepackage[utf8]{inputenc}
\usepackage[letterpaper,pdftex]{geometry}
\usepackage{longtable}
\usetikzlibrary{arrows,automata,fit,scopes,calc,matrix,positioning,decorations.pathmorphing,decorations.pathreplacing}
\pgfplotsset{width=9cm,compat=1.15}
\usepgfplotslibrary{statistics}
\usepackage[all]{xy}
\usepackage[affil-it]{authblk}
\usepackage[doublespacing]{setspace}
\usepackage{verbatim}
\usepackage{rotating}

\usepackage{hyperref}	
\hypersetup{colorlinks,linkcolor={blue},citecolor={blue!50!black},urlcolor={blue}}
\usepackage{geometry}

\title{Visitors Out! The Absence of Away Team Supporters as a Source of Home Advantage in Football}
\author{}
\author[1,2]{Federico Fioravanti \footnote{f.fioravanti@uva.nl (corresponding author)\\ Fioravanti acknowledges support from the Nederlandse Organisatie voor Wetenschappelijk Onderzoek Vici grant 639.023.811.}}
\author[2,3]{Fernando Delbianco} 
\author[2,3]{Fernando Tohm\'e}
\affil[1]{Institute for Logic, Language and Computation, University of Amsterdam, Netherlands}
\affil[2]{Instituto de Matemática (INMABB), Departamento de Matemática, Universidad Nacional del Sur (UNS)-CONICET, Bahía Blanca, Argentina}
\affil[3]{Departamento de Economía, Universidad Nacional del Sur (UNS)-Instituto de Matemática (INMABB)-CONICET, Bahía Blanca, Argentina}

\date{}

\begin{document}

\maketitle
\begin{abstract}
We seek to gain more insight into the effect of the crowds on the Home Advantage by analyzing the particular case of Argentinean football (also known as soccer), where for more than ten years, the visiting team fans were not allowed to attend the games. Additionally, during the COVID-$19$ lockdown, a significant number of games were played without both away and home team fans. The analysis of more than $20$ years of matches of the Argentinean tournament indicates that the absence of the away team crowds was beneficial for the Top $5$ teams during the first two years after their attendance was forbidden. An additional intriguing finding is that the lack of both crowds affects significantly all the teams, to the point of turning the home advantage into home `disadvantage' for most of the teams.
\textbf{JEL:} Z$20$
\end{abstract}
\section{Introduction}
The Home Advantage (HA) effect in sports has been thoroughly studied and confirmed, starting with the work of \cite{schwartz1977home}, considered to be the first empirical investigation of this effect. According to their findings, the presence of a local crowd has a stimulating effect, serving as a motivational force that enhances the performance of the home team. After their work, many other researchers investigated HA, with different approaches, taking into account considerations of different nature: physiological (\cite{neave2003testosterone}), psychological (\cite{agnew1994crowd}, \cite{legaz2013home}), economical (\cite{carmichael2005home},  \cite{boudreaux2017natural}, \cite{ponzo2018does}) and even exploring possible referee biases favoring home teams (\cite{downward2007effects}, \cite{page2010alone}, \cite{page2010evidence}, \cite{sacheti2015home}). These studies have been replicated for several sports. A vast literature can be found on HA in football (\cite{boyko2007referee}, \cite{taylor2008influence},  \cite{belchior2020fans}). But HA has been analyzed for other sports as well: rugby (\cite{thomas2008home}, \cite{garcia2013home}, \cite{dawson2020television}), basketball (\cite{jones2007home},  \cite{pojskic2011modelling}), tennis and golf (\cite{nevill1997identifying}), athletics (\cite{mccutcheon1984home}, \cite{jamieson2010home}), handball (\cite{aguilar2014determination}), judo (\cite{ferreira2013home}, \cite{krumer2017winning}). This phenomenon has been examined even in the context of major sports events such as the Olympic games and FIFA's World Cup (\cite{balmer2001home}, \cite{brown2002world}, \cite{balmer2003modelling}).\\
The COVID-$19$ pandemic gave a boost to the research on the HA effect since it forced almost all professional and amateur sports to be played without attending crowds. Thus, it induced a large-scale natural experiment on the impact of social pressure on decision-making and behavior in sports fields. Several analyses on its impact in football, such as \cite{bryson2021causal}, \cite{bilalic2021home}, \cite{mccarrick2021home}, and \cite{bilalic2023effect}, conclude that HA decreased during COVID-$19$ restricted tournaments (see \cite{leitner2022cauldron} for a literature review). This effect is also observed in other popular sports such as rugby union (\cite{delbianco2023home}), basketball (\cite{leota2022home}), ice hockey (\cite{thrane2023no}), and American Football (\cite{szabo2022impact}).  These results are consistent with the ones in \cite{pettersson2010behavior} and \cite{reade2022eliminating}, which considered games without attending crowds. Then one may conclude that HA can be detected even in the absence of drastic events like the COVID-$19$ lockdown. \cite{sedeaud2021covid} analyze rugby and football, finding that the HA effect is strongly dependent on the tournament, with the effect disappearing in the Premiership Rugby Championship\footnote{The English professional clubs' competition.}. \cite{ungureanu2021machine} use a complex dynamic systems approach, with machine learning tools to conclude that HA decreased and even disappeared in close games when compared to pre-COVID-$19$ seasons. On the contrary, a study by \cite{fazackerley2022influence} in rugby league, shows that players' performances are not affected by the presence or absence of crowds.\\
Although support from the fan base plays a role, a comprehensive understanding of this phenomenon requires the consideration of additional factors. The familiarity of the home team with the playing field (\cite{lago2011game}), the potential fatigue experienced by the away team due to travel (\cite{beckmann2022statistical}), and the social influence exerted by local fans over the referees officiating the game (\cite{zhang2022influence}); can have an impact on the HA effect. It is worth noting that the presence of an away crowd can sometimes counteract HA, introducing a new dynamic into the game. The support of the fans for the away team and their vocal presence can create a more competitive atmosphere, potentially adding an additional challenge to the performance of the home team. Hence, the interplay between the local and the visiting crowds becomes an essential aspect to weigh in the analysis of the influences on team performance in sports events. Thus, the conditions under which the Argentinean football tournament is played create an (almost) perfect experiment to test the influence of the visiting crowds. 
Argentinean football fans are world-known for their intensity and passion when supporting their respective teams.
They do not consider themselves as mere spectators but as the twelfth player. 
There is even a barra-brava\footnote{Latin-American version of the European ultras or English hooligans} from one of the most important teams in Argentina, Boca Juniors, that call themselves ``La $12$'' (``The 12'').
Smoke bombs, firecrackers, confetti, balloons, and the display of giant flags that cover entire stands, or part of them, before the match starts, are common scenarios that can be seen in Argentinean football stadiums.
But this behavior may lead, in the case of local tournaments, to clashes between fans, usually ending in violent riots. A death in a  clash in June of $2013$ forced the authorities to take radical action to prevent such incidents from happening again. Since then, the access of away team fans to the matches has been banned.\footnote{ See \href{https://www.bbc.com/news/world-latin-america-22864505}{BBC}, or \href{https://tn.com.ar/sociedad/2023/06/11/el-futbol-sin-hinchas-visitantes-cumple-10-anos-la-muerte-que-cambio-la-maxima-pasion-de-los-argentinos/}{TN} for more details in Argentinean media.}
This ban was originally thought to be held until the end of that tournament, but a later incident between Boca Juniors fans, that ended up with the death of two of them, made the authorities decide to hold the ban indefinitely. 
After the implementation of this policy, there were many (failed) isolated attempts to return to the ``normality''.
This ban will not likely be overturned in the foreseeable future, as the number of football-related deaths since the policy was implemented increased in comparison with the number of deaths during the previous $10$ years ($61$ deaths in $2003-2013$ and $71$ deaths in $2013-2023$).\footnote{See \href{https://www.telam.com.ar/notas/202306/630778-diez-anios-sin-hinchas-visitantes-futbol-argentino.html}{TELAM} (in Spanish).} This unusual setting allows us to compare the score difference and the winning rate of home teams (two simple ways of measuring the HA effect) with and without the attendance of the fans of the away team. The natural experiment that arose in $2020$ with the lockdown induced by the COVID-$19$ pandemic, where matches were played without any attendance, allows us to compare the previous results with the effects of playing without public at all. 
The analysis of the impact of bans on fans' attendance due to violent incidents, on the Home Advantage effect, has been considered in \cite{singleton2023decade}. They study the case of the Egyptian Premier League, which for almost a decade has been played without the public present at the stadiums. 
The period under analysis included the restriction policies applied during the COVID-$19$ pandemic.
Their results are consistent with the literature, where the HA decreased with the absence of fans, but there is no difference between the reasons for the absence (due to the ban or COVID-$19$ policies). \\
This study intends to quantify the influence of fans of both teams on the Home Advantage effect. We consider more than $20$ years of football matches played in the first level of Argentinean club competitions. Of those, ten years were played with mixed attendance, nine years with home crowds only, and one year with no public at all (due to COVID-$19$ restrictions). Thus, we have all the possible combinations other than the seemingly implausible case of games with only away team crowds in attendance. We find that the absence of away team fans has a positive effect for the five teams with the highest HA, but only for the first two years of this policy. We also find that the absence of local fans has a highly significant negative effect for all the teams of the tournament, making some teams to be subject to a `home disadvantage'.\\
Our results can be useful for sports planners wanting to increase a tournament's competitiveness. They can generate alternatives to disincentivize fans of the powerful teams to attend the games and incentivize fans of the less powerful teams to be at the matches. For example, they can offer top teams better broadcasting deals, so fans can watch the games on TV or the internet for a low price (or even for free) while subsidizing small clubs, lowering the tickets for their fans (or, again, giving them for free). Powerful teams enjoy a certain `slack': even with fewer home team fans in attendance, they manage to dominate most of the home games (but are still able to compensate for the box office loss with income from other sources). On the other hand, weaker teams cannot afford to lose their little home advantage due to their fan base and may have to exert an extra effort to induce the attendance of their fans (perhaps trading off more income for more frequent losses). \\
In this paper, we examine variables representing HA and see how they have been affected by the different restrictions imposed on the attendance of matches in the most important Argentinean football tournament for more than $10$ years. Sections $2$ and $3$ describe the database used in our study, and how some of the variables are defined. Section $4$ describes the model and presents the results. Finally, Section $5$ concludes and gives possible interpretations of the results.

\section{Data}
Our main data set consists of $7261$ football matches from all the seasons from $2003$ to $2022$ of the top Argentinean club's tournament. For commercial reasons, the tournament had different names during the period under examination. Besides these changes, since June $2013$, away crowds are no longer allowed into the stadiums. Thus $3389$ games were played with only home team fans in the stadium. The COVID-$19$ pandemic also forced (as in most parts of the world), after a relatively short suspension of the competition, to play $103$ games without public (`ghost games', as in \cite{leitner2021no}). That leaves us with $3769$ games played with no restriction on attendance. There is no available information about the number of spectators at each game. There are $44$ teams that played at least one tournament during the $2003-2022$ period. 
We use as a measure of the magnitude of Home Advantage the difference between the mean of goals scored by the home and the away team in a game. In Table \ref{tab:means} we can see that the HA during these $20$ years is approximately $0.32$ goals. We will analyze the evolution of the HA taking into account the presence/absence of the public, together with other factors that will be introduced in Section \ref{sec:empirical}. \\
\begin{table}[hbt!]
    \centering
    \begin{tabular}{c|c|c|c|c|c}
  Variable & Obs & Mean & Std. Dev. & Min & Max \\ \hline
Home goals & 7261 & 1.335629 & 1.150702 & 0 & 8 \\ 
Away goals & 7261 & 1.015014 & 1.036286 & 0 & 7 \\ 
Home Advantage & 7261 & 0.3203857 & 1.538001 & -6 & 7 \\ \hline
    \end{tabular}
    \caption{Descriptive Statistics}
    \label{tab:means}
\end{table}

\section{Motivation}\label{sec:motivation}

\begin{figure}[h]
    \centering
    \includegraphics[scale=0.55]{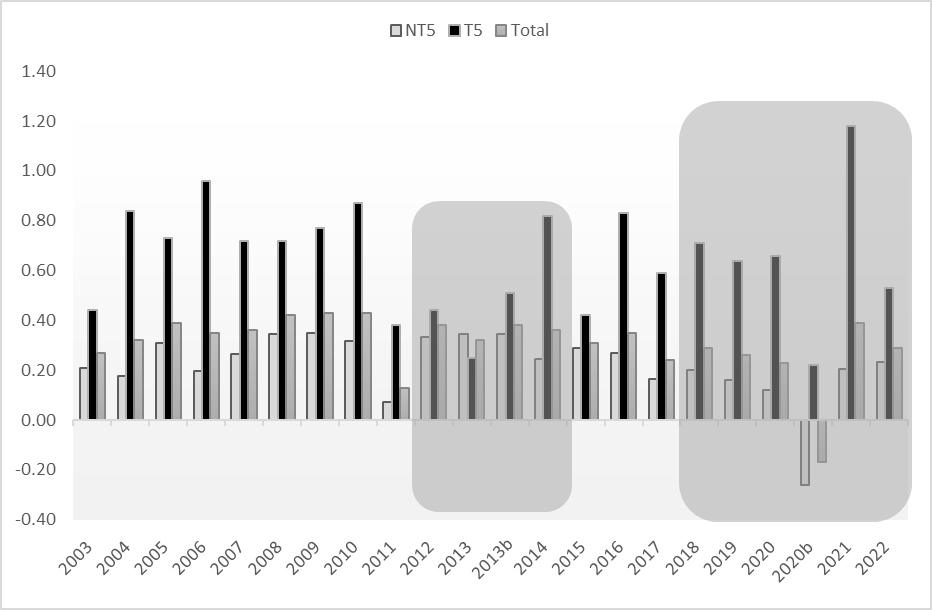}
    \caption{Distribution of annual HA by rank ({\em Total}: all the teams; {\em T5}: Top $5$; and {\em NT5}: Not-Top $5$ ones). Two intervals are highlighted, that will be the focus of a narrower analysis: $2012-2014$ and $2018-2022$.}
    \label{fig:graf1}
\end{figure}

\begin{figure}
    \centering
    \includegraphics[scale=0.85]{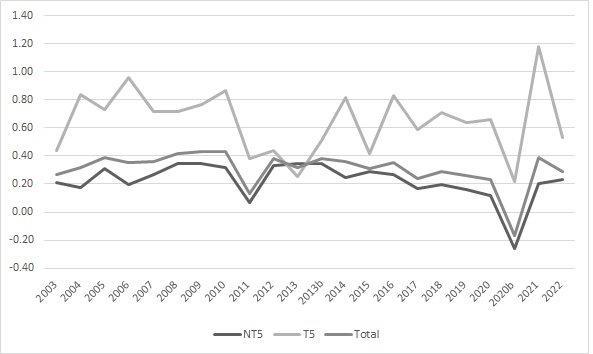}
    \caption{Trends in HA by rank.}
    \label{fig:line}
\end{figure}

Home Advantage is a complex phenomenon, that may have different causes. We are interested in isolating the effect of the attendance of visiting fans to the games using the natural experiment induced by the prohibition of visiting public in the matches of the main football tournament in Argentina.\\
We examine the hypothesis that teams that had a higher HA when attendance was free, have a larger HA when visitors were forbidden (see Table~\ref{tab:tab1} and Figure~\ref{fig:graf1}).\\
\begin{table}[h]
    \centering
    \begin{tabular}{c|c|c|c}
Year & Not-Top 5 & Top 5 & All teams \\ \hline
2003  & 0.21 & 0.44 & 0.27 \\
2004  & 0.18 & 0.84 & 0.32 \\
2005  & 0.31 & 0.73 & 0.39 \\
2006  & 0.20 & 0.96 & 0.35 \\
2007  & 0.27 & 0.72 & 0.36 \\
2008  & 0.35 & 0.72 & 0.42 \\
2009  & 0.35 & 0.77 & 0.43 \\
2010  & 0.32 & 0.87 & 0.43 \\
2011  & 0.07 & 0.38 & 0.13 \\
2012  & 0.33 & 0.44 & 0.38 \\
2013  & 0.34 & 0.25 & 0.32 \\
2013b & 0.35 & 0.51 & 0.38 \\
2014  & 0.25 & 0.82 & 0.36 \\
2015  & 0.29 & 0.42 & 0.31 \\
2016  & 0.27 & 0.83 & 0.35 \\
2017  & 0.16 & 0.59 & 0.24 \\
2018  & 0.20 & 0.71 & 0.29 \\
2019  & 0.16 & 0.64 & 0.26 \\
2020  & 0.12 & 0.66 & 0.23 \\
2020b  & -0.26 & 0.22 & -0.17 \\
2021  & 0.21 & 1.18 & 0.39 \\
2022  & 0.23 & 0.53 & 0.29 \\

    \end{tabular}
    \caption{Average HA by year and rank (all the teams, Top $5$ and Not-Top $5$ ones); $2013b$: Results after away teams fans are not allowed; $2020b$: Lockdown (no public allowed).}
    \label{tab:tab1}
\end{table}
\\
In particular, in Figure~\ref{fig:graf1} we can see the evolution of the HA during the years, for all the teams (Total column), and relative to the history of home performance of the teams (classified as Top $5$ and Not-Top $5$). Figure~\ref{fig:line} shows the evolution of HA in time, for the three groups of teams under analysis. The $5$ teams with the highest HA in the $2003-2022$ seasons are: {\it Boca Juniors}, {\it River Plate}, {\it Talleres de C\'ordoba}, {\it V\'elez Sarsfield}, and {\it Estudiantes de La Plata}. See Table \ref{tab:rankings} and Figure \ref{fig:histograma HA} in the Appendix for the complete ranking of the HA series by team. \\
We will focus, in particular, on the shaded areas of Figure~\ref{fig:graf1} which correspond to the periods in which new attendance policies were enacted. While in the long run, the teams may get used to playing with or without a home or away audience, in a narrow time window (one or two years after the new policy was implemented, for example) their performance may be affected by the different composition of the attending public. At first glance, this seems to be the case, especially in the $2018-2022$ shaded area.   \\

\section{Empirical Evidence}\label{sec:empirical}

We define the Home Advantage of a team in a game $i$, as
$$HA_i = GoalsH_i - GoalsA_i$$
\noindent where $GoalsH_i$ indicates the number of goals scored by the home team, while $GoalsA_i$ is the number of those scored by the away team. \\
We consider a categorical variable $Public_i$ with three values: $Free$, $Restricted$ (corresponding to games that can only be attended by home fans), and $Closed$ (matches played under lockdown with no public at all).\\
We consider a family of models $M_k$, defined as:
\begin{equation}
    HA_i = f_k(Public_i, HomeTeam_i, AwayTeam_i, Calendar_i)
\end{equation}
Each $f_k$ is a linear function, with a particular range of time during which the game was played ($Calendar_i$) and the restrictions enacted. We analyze the results for models $M_i$, described in Tables~\ref{tab:regressions},~\ref{tab:REG_ADD2},~\ref{tab:REG_ADD1}, and~\ref{tab:REG_ADD3}. In Table~\ref{tab:regressions}, we can see the results for the largest possible samples in each comparison. We do not consider teams' form variables, such as the dynamics ELO ratings or win percentages of teams, as they could become contaminated by the effects of playing in stadiums under different restrictions.\\
\begin{table}[]
    \centering
\begin{tabular}{lcccccc} \hline
 & M1 & M2 & M3 & M4 & M5 & M6 \\
VARIABLES & HA & HA & HA & HA & HA & HA \\ \hline
 &  &  &  &  &  &  \\
Free & 0.516*** & 0.0331 &  & 0.496*** & 0.513*** & 0.474*** \\
 & (0.154) & (0.0391) &  & (0.153) & (0.156) & (0.156) \\
Restricted & 0.487*** &  &  & 0.489*** & 0.447*** & 0.456*** \\
 & (0.154) &  &  & (0.153) & (0.154) & (0.157) \\
Closed &  &  & -0.487*** &  &  &  \\
 &  &  & (0.152) &  &  &  \\
Home team &  &  &  & YES &  &  \\
Away team &  &  &  &  & YES &  \\
Calendar &  &  &  &  &  & YES \\
Constant & -0.175 & 0.308*** & 0.312*** & -0.855*** & 0.298 & -0.121 \\
 & (0.151) & (0.0300) & (0.0255) & (0.212) & (0.221) & (0.166) \\
 &  &  &  &  &  &  \\
Observations & 7,261 & 6,383 & 3,661 & 7,261 & 7,261 & 7,261 \\
 R-squared & 0.002 & 0.000 & 0.003 & 0.032 & 0.037 & 0.005 \\
BP p-value & 0.987 & 0.277 & 0.718 & 0.668 & 0.020 & 0.881 \\\hline
\multicolumn{7}{c}{ Standard errors in parentheses. Robust standard errors for Model (5)} \\
\multicolumn{7}{c}{ *** p$<$0.01, ** p$<$0.05, * p$<$0.1} \\
\end{tabular}
    \caption{OLS regressions: $M1$: all teams both free and restricted entry against lockdown;  $M2$: free entry against restricted entry ($2003-2019$); $M3$: lockdown against restricted access ($2013-2022$); $M4$: all teams both free and restricted entry against lockdown plus home team dummy; $M5$: all teams both free and restricted entry against lockdown plus away team dummy; $M6$: all teams both free and restricted entry against lockdown plus tournament date dummy. BP denotes the results of the Breusch-Pagan heteroskedasticity test, displaying the p-value for the rejection of the null hypothesis of constant variance. The BP p-value for Model (5) corresponds to the OLS model without the later error correction.}
    \label{tab:regressions}
\end{table}\\

Another analysis involves a dependent variable $W_i$ that indicates whether the home team won game $i$ or not. That is, 
$W_i=1$ if $HA_i > 0$ and $W_i = 0$ otherwise. This allows us to examine the impact of the restrictions on the access to the games on the probability of winning, instead of on the difference in scores. That is, each model $L_k$ is
\begin{equation}
  W_i = g_k(Public_i)  
\end{equation}
where each $g_k$ is defined by the number of teams considered and the period in which the games were played (see Tables~\ref{tab:logits2},~\ref{tab:logits4}, and~\ref{tab:logits3}).\\

\begin{table}[h!]
    \centering
    \begin{tabular}{lcccccc} \hline
 & L1 & L2 & L3 & L4 & L5 & L6 \\
VARIABLES & W & W & W & W & W & W \\ \hline
 &  &  &  &  &  &  \\
Free & 0.455** & -0.0273 &  & 0.461** & 0.455** & 0.462** \\
 & (0.212) & (0.0513) &  & (0.216) & (0.216) & (0.216) \\
Restricted & 0.475** &  &  & 0.440** & 0.505** & 0.484** \\
 & (0.212) &  &  & (0.215) & (0.217) & (0.215) \\
Closed &  &  & -0.469** &  &  &  \\
&  &  & (0.212) &  &  &  \\
Home Team &  &  &  & YES &  &  \\
Away Team &  &  &  &  & YES &  \\
Calendar &  &  &  &  &  & YES \\
Constant & -0.708*** & -0.225*** & -0.239*** & -0.309*** & -1.217 & -0.717*** \\
 & (0.210) & (0.0394) & (0.0338) & (0.291) & (0.301) & (0.228) \\
 &  &  &  &  &  &  \\
 Observations & 7,261 & 6,383 & 3,661 & 7,261 & 7,261 & 7,261 \\ \hline
\multicolumn{7}{c}{ Standard errors in parentheses} \\
\multicolumn{7}{c}{ *** p$<$0.01, ** p$<$0.05, * p$<$0.1} \\
\end{tabular}
    \caption{Logit regressions following OLS regressions controls of Table \ref{tab:regressions}}
    \label{tab:logits2}
\end{table}

We split our analysis to consider the two main events highlighted in Figure~\ref{fig:graf1}, namely the prohibition of attendance of visitor fans and the COVID-$19$ lockdown. We start the analysis in chronological order.\\
Some clarifications of the statistical work carried out are in order. We define the Top $5$ as the teams that during the $20$ years covered by our sample ranked on average as one of the five highest values of HA. The Not-Top $5$, consequently, are the remaining $39$ teams. It is worth noticing that these Top $5$ teams are not necessarily the ``historical'' Top $5$ (also known as ``Big Five''), the most successful teams in Argentinean history, namely River, Boca Juniors, Racing, Independiente, and San Lorenzo. There are many reasons for not using the ``Big Five''. First, we found that there is no significant effect by considering this group of teams. A possible reason for these clubs not being dominant on the HA effect is that over the last $20$ years, there are other teams that were more successful. Finally, as we want to consider how the HA changed under different conditions on attendance, we wanted to compare the best teams regarding this characteristic, with the rest of the sampled teams.  \\
Local derbies have been considered to have an effect on the HA (see, for example, \cite{volossovitch2013home}), but we find that there is not a significant effect of it on our sample, which is why we disregard this variable for this study (see Table~\ref{tab:REG_ADD3} in the Appendix).\\
On the other hand, we consider different temporal subsamples. First the total sample. Then, we will consider $2003-2019$, comparing the crowds of both teams vs. just that of the local team. Thirdly, we use $2013-2022$ to contrast only home public against a closed stadium. Finally, we will focus on the periods close to the policy changes, $2012-2014$ and $2018-2022$ on the other.\\
To assess the regression results, we carried out heteroscedasticity tests controlling for indicator variables that capture the heterogeneity between teams. Only in one of the specifications does the Breusch-Pagan test reject the null hypothesis of homoscedasticity. In this case, a robust White estimate yields similar results in the standard deviations and maintains the significance of the variables (the standard deviations change a few hundredths, to $0.155$ and $0.156$ for the {\it Free} and {\it Restricted} contexts, respectively).

\subsection{Visitors Out}

\begin{figure}[h]
    \centering
    \includegraphics[scale=0.60]{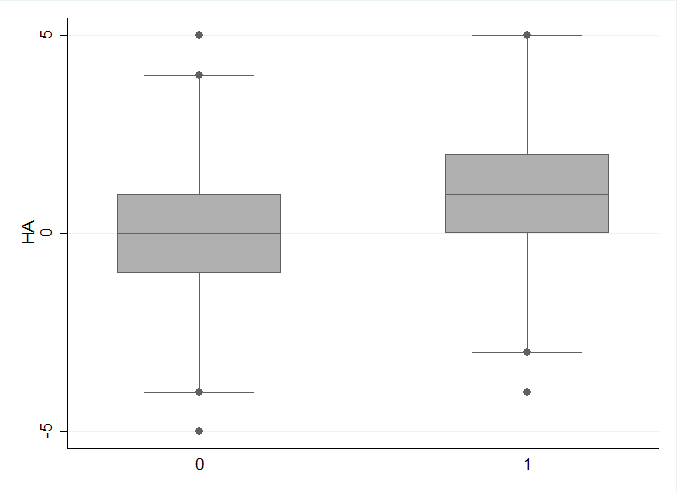}
    \caption{Home Advantage (HA) Box-and-Whiskers plot of the averages of Not-Top 5 (0) versus Top 5 teams (1) for seasons $2012-2014$.}
    \label{fig:box}
\end{figure}

First, we focus now on the seasons after $2013$ and before $2019$ in which only visiting fans were banned. That is when teams played only with the local supporting public.\\
The full sample does not exhibit significant effects on the goal difference in favor of the home team, nor on the probability of winning of teams playing locally. This can be seen in both linear and logistic regressions (see Table~\ref{tab:regressions} and Table~\ref{tab:logits2}). But if we restrict the analysis to the Top $5$, comparing it to the rest of the teams, we can see in Figure~\ref{fig:graf1} and Table~\ref{tab:REG_ADD2} that its average increases (but not significatively) both over the second half of $2013$ (when the restriction began), and in $2014$ (and even more during the second year of the restriction). But from $2015$ to $2019$ their HA went back to its pre-restriction historical values.
Besides not being a highly significant effect, even for the period close to the policy´s implementation, a possible explanation is that in the long run, teams get used to playing only with local fans.\\
\begin{table}[h!]
    \centering
\begin{tabular}{lcccccc} \hline
 & M7 & M8 & M9 & M10 & M11 & M12\\
VARIABLES & HA & HA & HA & HA & HA & HA\\ \hline
 &  &  &  &  \\
Free & 0.0331 & 0.0271 & -0.0654 & -0.329  & -0.366 & -0.112\\
 & (0.0391) & (0.0909) & (0.0898) & (0.202) & (0.233) & (0.268)\\
Home Team  &  &  &  &  & YES & YES\\
Away Team &  &  &  &  & YES & YES \\
Calendar &  &  &  &  & YES & YES \\
Constant & 0.308*** & 0.659*** & 0.367*** & 0.709*** & 0.917 & 1.097***\\
 & (0.0300) & (0.0719) & (0.0627) & (0.137) & (0.789) & (0.368)\\
 &  &  &  & &  &  \\
Observations & 6,383 & 1,208 & 1,130 & 217 & 217 & 1130\\
 R-squared & 0.000 & 0.000 & 0.000 & 0.012 & 0.170 & 0.093\\ \hline
\multicolumn{7}{c}{ Standard errors in parentheses} \\
\multicolumn{7}{c}{ *** p$<$0.01, ** p$<$0.05, * p$<$0.1} \\
\end{tabular}
    \caption{M7: all teams $2003-2019$; M8: Top $5$ teams, $2003-2019$; M9: all teams, $2012-2014$; M10: Top $5$ teams, $2012-2014$; M11: Top $5$ teams, $2012-2014$ with dummies; M12: all teams, $2012-2014$ with dummies}
    \label{tab:REG_ADD2}
\end{table}\\
When we consider the winning percentages, the results are consistent with the ones in Table~\ref{tab:REG_ADD2}, but become significant ($p<0.1$) for the Not-Top $5$ teams (see Table~\ref{tab:logits4} in the Appendix). These teams, although not consistently scoring more goals, won more games when playing at home. And intuitive reasoning behind this result is that ``small'' teams benefited when playing in their stadium without the ``overwhelming'' crowds of the bigger teams; and won matches that they previously draw or lost by a small margin when attendance was free.\\
If we run a difference of means test to compare the average HA for the Top $5$ teams before and after the no-visitor's policy was implemented, we find that a positive difference in their HA is significant only at $10\%$ with a p-value of $0.052$, as can be seen in Table \ref{tab:mean_testsNV}. This is due to the great dispersion of results in the different games that were played. It is interesting to note that this result, found for the Top $5$, does not hold for the Top $2$ ({\it Boca Juniors} and {\it River Plate}, arguably the most popular teams in Argentina) or for the Top $10$ teams. In the former case, {\it Boca Juniors} and {\it River Plate} are the teams with the highest HA (approximately one goal of difference over visiting teams) and are hardly affected by the presence or absence of visiting fans. In the case of Top $10$ teams, the averages of the teams in the $6-10$ range are not far from the rest of the Not-Top $5$ teams. These results are shown in the Table~\ref{tab:mean_testsNV}. We can see that the averages are, respectively $0.55$ and $0.52$ and thus their difference is very small.
\begin{table}[h!]
    \centering
    \begin{tabular}{c|c|c|c}
Sample & Top5 & Top2 & Top10\\
Mean Free & 0.38 & 0.68 & 0.55\\
Mean Restricted & 0.71 & 0.79 & 0.52\\
$Pr(T<t)$ & 0.948 & 0.6388 & 0.4184\\
$Pr(T>t)$ & 0.052 & 0.3612 & 0.5816\\ \hline 
    \end{tabular}
    \caption{Difference of means test for the period $2012-2014$}
    \label{tab:mean_testsNV}
\end{table}\\
It can be seen in Figure \ref{fig:box} that the means for Top $5$ teams are higher (the box corresponding to the Top $5$ is higher), but the dispersion almost overlaps the intervals (the whiskers of the boxes reach similar intervals, with a few outliers).\\

\subsection{COVID-19}

Columns $1$ and $3$ of Tables~\ref{tab:regressions} and \ref{tab:logits2} show that the lockdown affected the advantage of home teams, compared to the periods in which attendance was free as well as when it was forbidden just for visiting fans. This result is consistent with the literature cited in Section $1$ on the effects of the lockdown in football and other team sports. This result is robust to the presence of several dummies controlling for the tournament calendar ({\it Calendar}) and fixed effects of teams (Home and Away teams). \\
When we narrow down to focus on the Top $5$ and Not Top $5$, in Table~\ref{tab:REG_ADD1} we can see that there is a negative but not significant effect on the HA for the Top $5$ teams and a highly significant negative effect for the Not-Top $5$ teams.\\
If we consider the effect of the lockdown on the winning percentage (see Table~\ref{tab:logits3} in the Appendix), we can see that the effect is still significant, but it loses significance for the Not-Top $5$ teams. This possibly implies that home teams scored less, but still got the previous results (now obtaining narrow victories or narrow defeats).\\
When considering the difference of means, the decrease in the HA becomes significant ($p<0.1$) for the Top $5$, while this negative effect is highly significant for Not-Top $5$, Not-Top $2$ and Not-Top $10$ (see Table~\ref{tab:mean_testsCOVID}).\\
\begin{table}[h!]
    \centering
\begin{tabular}{lccccc} \hline
 & M13 & M14 & M15 & M16 & M17 \\
VARIABLES & HA & HA & HA & HA & HA \\ \hline
 &  &  &  &  &  \\
Closed & -0.502*** & -0.476*** & -0.525 & -0.455*** & -0.433*** \\
 & (0.153) & (0.156) & (0.367) & (0.170) & (0.153) \\
Home Team &  &  &  &  & YES \\
Away Team &  &  &  &  & YES \\
Calendar &  &  &  &  & YES \\
Constant & 0.328*** & 0.301*** & 0.747*** & 0.196*** & 0.00595 \\
 & (0.0182) & (0.0395) & (0.0894) & (0.0435) & (0.235) \\
 &  &  &  &  &  \\
Observations & 7,261 & 1,605 & 303 & 1,302 & 7,261 \\
 R-squared & 0.001 & 0.006 & 0.007 & 0.005 & 0.070 \\ \hline
\multicolumn{6}{c}{ Standard errors in parentheses} \\
\multicolumn{6}{c}{ *** p$<$0.01, ** p$<$0.05, * p$<$0.1} \\
\end{tabular}
    \caption{M13: all teams, free and restricted vs closed, $2003-2022$; M14: all teams, restricted vs closed, $2018-2022$; M15: Top $5$, restricted vs closed, $2018-2022$; M16: Not-Top $5$, restricted vs closed, $2018-2022$; M17: all teams, free and restricted vs closed, $2003-2022$ adding dummies. }
    \label{tab:REG_ADD1}
\end{table}
\begin{table}[h!]
    \centering
    \begin{tabular}{c|c|c|c}
 \hline
Sample & Top5 & Top2 & Top10\\
Mean Restricted & 0.74 & 1 & 0.59\\
Mean Closed & 0.22 & 0.57 & 0\\
$Pr(T<t)$ & 0.9235 & 0.7477 & 0.9906\\
$Pr(T>t)$ & 0.075 & 0.2523 & 0.0094\\ \hline \hline
Sample & NotTop5 & NotTop2 & NotTop10\\
Mean Restricted & 0.19 & 0.24 & 0.15\\
Mean Closed & -0.25 & -0.22 & -0.27\\
$Pr(T<t)$ & 0.9962 & 0.9985 & 0.9793\\
$Pr(T>t)$ & 0.0038 & 0.0015 & 0.0207\\ \hline
    \end{tabular}
    \caption{Difference of means test for the period $2018-2022$}
    \label{tab:mean_testsCOVID}
\end{table}
Three results can be deduced from the regressions as well as from the descriptive statistics. The first is that in fact, the vast majority of teams exhibit a ``negative'' HA. This can be seen in the total averages by year and in the HA rankings. All the teams not in the Top $5$ were harmed by not playing with their public. On the other hand, the Top $5$ teams kept a certain degree of advantage, but lower than their historical advantage (this can be seen in the black bar of the annual histogram). One possible interpretation of this is that the HA effect for teams that are not in the Top $5$ depends only on the presence or absence of local fans. This means that a better knowledge of the field or the fatigue of the visiting team, are not significant factors for the HA effect. The best teams (those in the Top $5$), require further analysis to disentangle all the factors influencing the HA effect, although our results show that the presence of local fans has a major impact.\\
The second realization arose immediately after the return of the local public, once the COVID-$19$ lockdown was lifted (remember that visiting fans are still banned in local championships). During that period the Top $5$ teams achieved a historical peak of average home goal advantage. This could be interpreted as an extra boost generated by local fans. The rest of the teams recovered well, on average, but did not perform much better than their historical record. This difference may ensue from the fact that most powerful teams have a larger number of fans. These teams transitioned in this period from playing with empty (generally big) stadiums to playing with a full big stadium filled with their own fans.\\
The last piece of evidence is that the effects of the lockdown are significant compared to both the $2003-2012$ season, as well as the $2013-2019$ and $2021-2022$ seasons. In other words, the HA decreases to a considerable extent in comparison to both the case in which fans of both teams attend and the case with only the local public.
\section{Conclusions}

In this work, we studied the effect of the attendance of crowds (and in particular of visiting fans) on the Home Advantage in the context of the most important football club tournament in Argentina. The ban on visitors after $2013$, due to a violent incident, and the COVID-$19$ lockdown, forcing the total absence of public, created two natural experiments that allow us to investigate the effect of the presence of home and away fans separately. Despite the Home Advantage effect has been established in the literature, it is not easy to find situations in which some of its causes can be isolated. This paper intends to identify whether the attendance of home and away crowds have an impact on the outcome of a football game, at least in Argentina.\\
When we compare the mean score difference and the winning rates of games with full attendance and games with only local fans, we do not find significant differences. For the Top $5$ teams, an increase in the HA can be detected in the first two years of the implementation of the `no visitors' policy. A possible interpretation is that smaller teams, when playing as visitors against the Top $5$ teams, were `extremely' demotivated by the chanting of the local fans and got used to that only two years later. It is worth mentioning that this effect is no longer significant when we consider the Top $2$ teams, {\it Boca Juniors} and {\it River Plate}. This is not so surprising, as they enjoy a significantly large HA, and it is natural to assume that the presence of visiting fans might not have a large effect.\\
We find that the HA significantly decreased for all the teams during the COVID-$19$ restrictions, when no public was allowed in the stadiums. This is consistent for HA considered as score differences and winning rates. It is important to notice that although this effect was important for all teams, the Top $5$ teams still kept a positive HA, while for the rest of the teams, the HA became negative. One possible interpretation is that factors other than the presence of the home crowd may not be significant for the HA of weaker teams. For top teams, although it has a major effect, further research is needed to assess the importance of other factors. Besides that, the return of local fans gave a further boost, for a year, to the Top $5$ teams. This can be interpreted as indicating that for the stronger teams, with larger stadiums and larger numbers of fans, the transition from playing in empty stadiums to playing with their fans in attendance is more beneficial than for teams with smaller fan bases. Although the effects of the lockdown are significant for both measures of the HA, it is stronger for the score difference of the Not-Top $5$ teams. One can possibly infer from this that home teams scored less (diminishing the difference), but, on average, kept winning and losing the same number of games, now obtaining tighter results.\\
The natural experiments generated by Argentinean football after the ban on visitor fans and during the COVID-$19$ pandemic allow us to have a closer look at the Home Advantage effect, which has been widely studied and confirmed in the literature. The two prohibitions created the context in which we could disentangle the way local and away crowds affect HA. While the impact of the presence of local fans is significant, the absence of visitors is important only for a few teams. Further research is needed to unravel how crowd attendance might affect home-field advantage, whether by exerting social pressure on referees or by cheering up the home team (or even demotivating the away team).

\bibliography{ref}

\newpage
\section*{Appendix}

\begin{table}[hbt!]
    \centering
    \begin{tabular}{|c|c|c|c|}
        \hline
Home & HA average & Away & HA average (opponent) \\ \hline
Huracán (TA) & -0.736 & River Plate & -0.199 \\ 
Tiro Federal & -0.736 & Boca Juniors & -0.196 \\ 
Crucero del Norte & -0.4666 & Vélez Sarsfield & -0.012 \\ 
Almagro & -0.444 & Defensa y Justicia & 0.043 \\ 
Aldosivi & -0.397 & San Lorenzo & 0.068 \\ 
Nueva Chicago & -0.301 & Estudiantes & 0.105 \\ 
Instituto & -0.289 & Talleres (C) & 0.161 \\ 
Chacarita Juniors & -0.254 & Independiente & 0.161 \\ 
Sarmiento & -0.222 & Racing Club & 0.178 \\ 
Central Córdoba (SdE) & -0.170 & Lanús & 0.185 \\ 
Temperley & -0.115 & Banfield & 0.185 \\ 
Quilmes & -0.033 & Newells & 0.252 \\ 
Atlético de Rafaela & 0.024 & Belgrano & 0.335 \\ 
San Martín (T) & 0.031 & Argentinos & 0.352 \\ 
Gimnasia y Esgrima (J) & 0.093 & Platense & 0.360 \\ 
Unión & 0.117 & Tigre & 0.369 \\ 
Olimpo & 0.125 & Godoy Cruz & 0.373 \\ 
San Martín (SJ) & 0.132 & Rosario Central & 0.400 \\ 
Patronato & 0.133 & Arsenal & 0.440 \\ 
Arsenal & 0.187 & Unión & 0.444 \\ 
Colón & 0.188 & Colón & 0.448 \\ 
Gimnasia y Esgrima (LP) & 0.205 & Central Córdoba (SdE) & 0.474 \\ 
All Boys & 0.226 & Gimnasia y Esgrima (LP) & 0.513 \\ 
Banfield & 0.228 & Atlético Tucumán & 0.527 \\ 
Argentinos & 0.230 & Huracán & 0.560 \\ 
Huracán & 0.250 & Atlético de Rafaela & 0.581 \\ 
Platense & 0.269 & Sarmiento & 0.593 \\ 
Godoy Cruz & 0.275 & San Martín (SJ) & 0.641 \\ 
Barracas Central & 0.285 & San Martín (T) & 0.645 \\ 
Tigre & 0.287 & Chacarita Juniors & 0.653 \\ 
Belgrano & 0.295 & Quilmes & 0.661 \\ 
Rosario Central & 0.310 & Gimnasia y Esgrima (J) & 0.684 \\ 
Atlético Tucumán & 0.378 & Aldosivi & 0.720 \\ 
Independiente & 0.436 & All Boys & 0.723 \\ 
San Lorenzo & 0.438 & Patronato & 0.739 \\ 
Defensa y Justicia & 0.443 & Nueva Chicago & 0.754 \\ 
Lanús & 0.453 & Barracas Central & 0.769 \\ 
Racing Club & 0.476 & Olimpo & 0.773 \\ 
Newells & 0.487 & Almagro & 0.789 \\ 
Estudiantes & 0.540 & Temperley & 0.843 \\ 
Vélez Sarsfield & 0.562 & Tiro Federal & 1.000 \\ 
Talleres (C) & 0.584 & Instituto & 1.000 \\ 
River Plate & 0.737 & Huracán (TA) & 1.631 \\ 
Boca Juniors & 0.932 & Crucero del Norte & 1.800 \\ \hline

    \end{tabular}
    \caption{Ranking of average HA results}
    \label{tab:rankings}
\end{table}

\begin{figure}[!ht]
    \centering
    \includegraphics[scale=0.6]{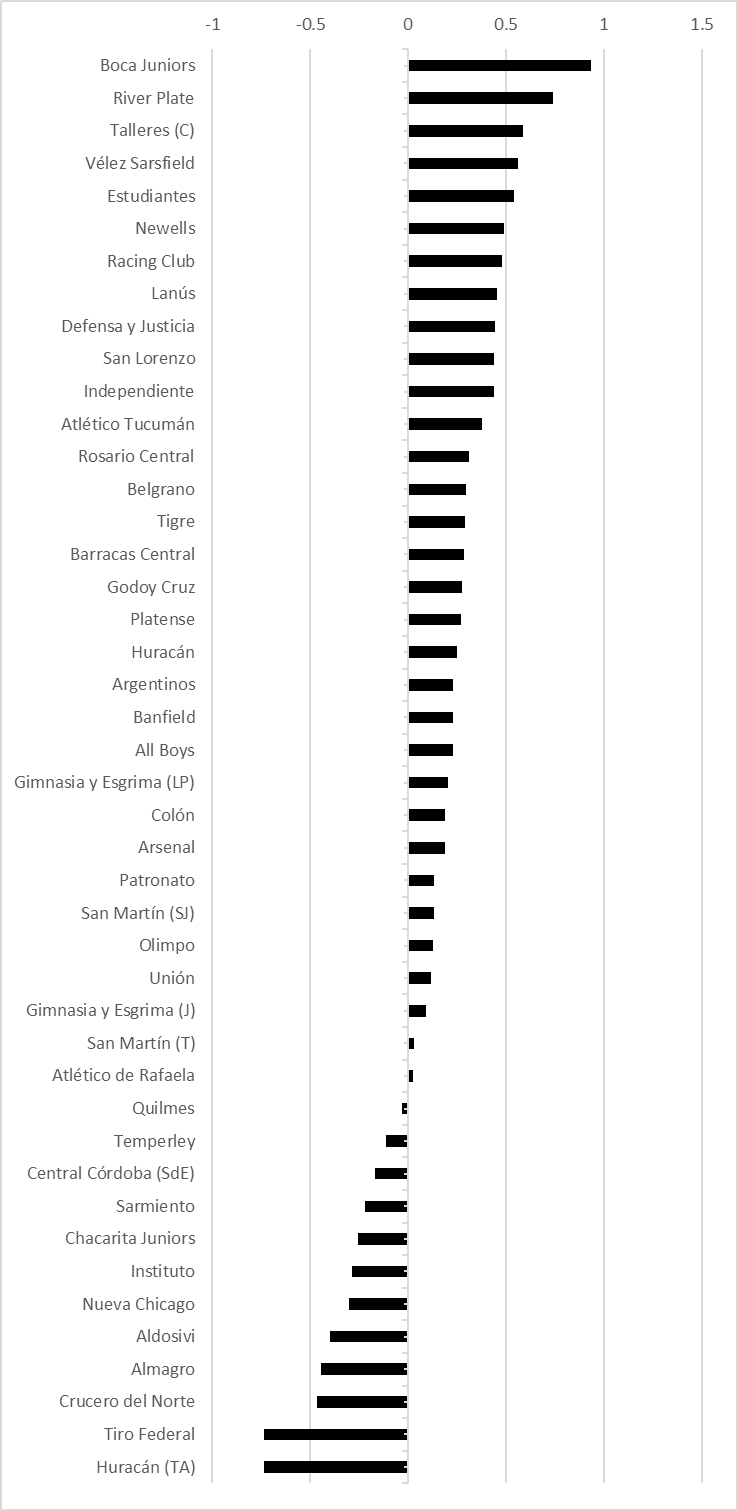}
    \caption{Distribution of the HA for the $44$ teams on the period $2003-2022$}
    \label{fig:histograma HA}
\end{figure}

\begin{table}[h]
    \centering
    \begin{tabular}{lcc} \hline
 & M18 & L18  \\
VARIABLES & HA & W \\ \hline
 &  &   \\
Free & 0.516*** & 0.108**  \\
 & (0.154) & (0.0496)  \\
Restricted & 0.486*** & 0.113**  \\
 & (0.154) & (0.0496)  \\
Derby & 0.0241 & -0.0323  \\
& (0.118) & (0.0381)  \\
Constant & -0.175 & 0.330***  \\
 & (0.151) & (0.0489)  \\
 &  &   \\
Observations & 7,260 & 7,261  \\
 R-squared & 0.002 & 0.001  \\ \hline
\multicolumn{3}{c}{ Standard errors in parentheses} \\
\multicolumn{3}{c}{ *** p$<$0.01, ** p$<$0.05, * p$<$0.1} \\
\end{tabular}
    \caption{OLS and logit regressions of the full sample (2003-2022) and all teams, with a dummy for derby matches.}
    \label{tab:REG_ADD3}
\end{table}
\begin{table}[h]
    \centering
\begin{tabular}{lcccccc} \hline
 & L7 & L8 & L9 & L10 & L11 & L12\\
VARIABLES & W & W & W & W & W & W\\ \hline
 &  &  &  &  &  & \\
Free & -0.0273 & -0.0207 & -0.150 & -0.488* & -0.756** & -0.231 \\
 & (0.0513) & (0.119) & (0.120) & (0.275) & (0.352) & (0.144)\\
Home Team  &  &  &  &  & YES & YES\\
Away Team &  &  &  &  & YES & YES \\
Calendar &  &  &  &  & YES & YES \\
Constant & -0.225*** & 0.186** & -0.173** & 0.328* & 0.584 & 0.431 \\
 & (0.0394) & (0.0943) & (0.0834) & (0.187) & (1.137) & (0.521)\\
 &  &  &  &  &  &\\
 Observations & 6,383 & 1,208 & 1,130 & 217 & 217 & 1130 \\ \hline
\multicolumn{7}{c}{ Standard errors in parentheses} \\
\multicolumn{7}{c}{ *** p$<$0.01, ** p$<$0.05, * p$<$0.1} \\
\end{tabular}
    \caption{Logits regressions of models in Table \ref{tab:REG_ADD2}}
    \label{tab:logits4}
\end{table}

\begin{table}[h]
    \centering
    \begin{tabular}{lccccc} \hline
 & L13 & L14 & L15 & L16 & L17 \\
VARIABLES & W & W & W & W & W \\ \hline
 &  &  &  &  &  \\
Closed & -0.465** & -0.448** & -0.670 & -0.391 & -0.476** \\
 & (0.211) & (0.216) & (0.498) & (0.240) & (0.221) \\
Home Team &  &  &  &  & YES \\
Away Team &  &  &  &  & YES \\
Calendar &  &  &  &  & YES \\
Constant & -0.243*** & -0.260*** & 0.218* & -0.374*** & -0.445 \\
 & (0.0238) & (0.0520) & (0.119) & (0.0583) & (0.332) \\
 &  &  &  &  &  \\
 Observations & 7,261 & 1,605 & 303 & 1,302 & 7,261 \\ \hline
\multicolumn{6}{c}{ Standard errors in parentheses} \\
\multicolumn{6}{c}{ *** p$<$0.01, ** p$<$0.05, * p$<$0.1} \\
\end{tabular}
    \caption{Logits regressions of models in Table \ref{tab:REG_ADD1}}
    \label{tab:logits3}
\end{table}
\end{document}